\documentclass[aps,pre,groupedaddress,superscriptaddress,showkeys,showpacs,amsmath]{revtex4-1}
\usepackage{amssymb}
\usepackage{amsmath}
\usepackage{graphicx}
\usepackage{amsbsy}
\usepackage{ulem}
\usepackage{bm}
\usepackage{color}
\usepackage{gensymb}
\usepackage{caption}
\usepackage{subcaption}

\begin{document}

\title{The elixir phase of chain molecules}

\author{Tatjana \v{S}krbi\'{c}} 
\email{tatjana.skrbic@unive.it}
\affiliation{Dipartimento di Scienze Molecolari e Nanosistemi, 
Universit\`{a} Ca' Foscari di Venezia
Campus Scientifico, Edificio Alfa,
via Torino 155,30170 Venezia Mestre, Italy}

\author{Trinh X. Hoang}
\email{hoang@iop.vast.ac.vn}
\affiliation{Center for Computational Physics, Institute of Physics, Vietnam
Academy of Science and Technology, 10 Dao Tan, Ba Dinh, Hanoi, Vietnam}

\author{Amos Maritan}
\email{amos.maritan@pd.infn.it}
\affiliation{Dipartimento di Fisica e Astronomia, Universit\`a di Padova, and INFN, via Marzolo 8,I-35131 Padova, Italy}

\author{Jayanth R. Banavar}
\email{banavar@uoregon.edu}
\affiliation{Department of Physics and Institute for Theoretical Science, 1274 University of Oregon, 
Eugene, OR 97403-1274, USA}

\author{Achille Giacometti} 
\email{achille.giacometti@unive.it}
\affiliation{Dipartimento di Scienze Molecolari e Nanosistemi, 
Universit\`{a} Ca' Foscari di Venezia
Campus Scientifico, Edificio Alfa,
via Torino 155,30170 Venezia Mestre, Italy}

\date{\today}

\begin{abstract}
A phase of matter is a familiar notion for inanimate physical matter.  The nature of a phase of matter transcends the microscopic material properties. For example, materials in the liquid phase have certain common properties independent of the chemistry of the constituents: liquids take the shape of the container; they flow; and they can be poured -- alcohol, oil and water as well as a Lennard-Jones computer model exhibit similar behavior when poised in the liquid phase. Here we identify a hitherto unstudied `phase' of matter, the elixir phase, in a simple model of a polymeric chain whose backbone has the correct local cylindrical symmetry induced by the tangent to the chain. The elixir phase appears on breaking the cylindrical symmetry by adding side spheres along the negative normal direction, as in proteins. This phase, nestled between other phases, has multiple ground states made up of building blocks of helices and almost planar sheets akin to protein native folds. We discuss the similarities of this `phase' of a finite size system to the liquid crystal   and spin glass phases. Our findings are relevant for understanding proteins; the creation of novel bio-inspired nano-machines; and also may have implications for life elsewhere in the cosmos.
\end{abstract}

\maketitle
\section{Introduction}
\label{sec:introduction}

Polymer science \cite{Flory69,deGennes79,Khokhlov02,Rubinstein03}, the study of chain molecules (linear polymer), is a flourishing subject that has led to path-breaking advances in plastics, textiles, and a variety of other technologies. There are simple, yet powerful, paradigms for understanding chain molecules. A polymer chain is composed of many interacting monomers that are tethered together to form a linear array. If the only interaction is excluded volume or self-avoidance, a single chain is in a self-avoiding coil phase whose large scale behavior is distinct from that of a fully non-interacting chain in both $2$ and $3$ spatial dimensions. Upon adding self-attraction between pairs of non-adjacent monomers promoting chain compaction, there is a phase transition from a random coil phase at high temperatures to a highly degenerate compact phase at low temperatures \cite{Flory69,deGennes79,Khokhlov02}. Note that the notion of phases and phase transitions for a polymeric chain at non-zero temperatures strictly refer to a chain with an infinite number of monomers. Indeed it is only in that limit that singularities in the thermodynamic potentials arise. However real systems, and the polymer chain is no exception, are never infinite. Thus we expect that, for long enough chains, we can identify the {\it memory} of the phases and the associated phase transitions in the infinite size limit. For finite size systems including an interacting many-body linear chain of finite length, phases as well as transitions between them can and do exist at zero temperature as some parameters of the system are varied. The existence of a first order phase transition at zero temperature is typically signaled by the existence of metastable states at non-zero temperature that can persist for long time scales depending on the temperature and on the chain length. This is the context within which we carry out our analysis of a model of a protein, a finite length bio-polymer.

A globular protein is a hetero-polymer - a linear chain whose monomers can be of twenty types of naturally occurring amino acids with distinct side chains \cite{Creighton92}. Thus the number of possible sequences of a protein just 40 amino acids long is $20^{40}$, an astronomically large number. Not all these sequences exhibit archetypical protein like behavior  -- folding rapidly and reproducibly under physiological conditions to their native state structures \cite{Anfinsen73}. This is typically described in terms of two phases, the unfolded and folded, separated by a first order transition, with the implicit meaning described above. Interestingly, protein structures are frequently tolerant to changes in amino acid sequence with mutations often causing only modest local changes in structure \cite{Baase10}. Remarkably, the total number of distinct protein native state folds is limited and only of the order of a few thousand \cite{Chothia92}.  This follows from the fact that native state folds are made of building blocks of tightly curled helices and strands assembled into almost planar sheet like motifs. The number of distinct topologies in which these motifs can be assembled is limited \cite{Chothia77,Przytycka99}. The existence of these secondary motifs follows from two independent considerations: scaffolding provided by hydrogen bonds \cite{Pauling51,Pauling51b}; and the steric requirement to avoid atomic overlaps \cite{Ramachandran68}.

Flory \cite{Flory69} wrote in 1969, {\it Synthetic analogs of globular proteins are unknown. The capability of adopting a dense globular configuration stabilized by self-interactions and of transforming reversibly to the random coil are characteristics peculiar to the chain molecules of globular proteins alone.} We note however that modern chemistry has made the engineering of synthetic analogs of globular proteins possible.
The model proposed here provides a route for creating synthetic analogs of globular proteins and provides a bridge between polymer science and bio-molecular science.

We begin with a standard classic homo-polymer model. The linear chain comprises $N$ main chain spheres of diameter $\sigma$ in a railway train topology with, for simplicity, the distances between the centers of successive monomers set to be exactly $b$.  In order to avoid spurious symmetries, we allow for the overlap of adjoining spheres ($\sigma \geq b$). Symmetry plays a key role in determining the phases of matter. For example, sensitive liquid crystal phases\cite{Chaikin00} form when a collection of spheres is replaced by a collection of anisotropic objects such as pencils or ellipsoids. The liquid crystal phase occurs at non-zero temperatures close to the liquid phase. 

A sphere is isotropic and looks the same when viewed from any direction. In contrast, there is a preferred axis at the location of each main chain sphere corresponding to the tangent along the chain or the direction along which the chain is oriented at that location. Allowing adjoining spheres to overlap breaks the spherical symmetry of an individual main chain sphere. The overlap between adjoining main-chain spheres results in an entropic stiffness because of the reduction in the ability to bend the chain \cite{Skrbic16b}. A hint that such symmetry breaking could be important is provided by the extreme case of a chain of coins, which, in the continuum limit, has a tube-like geometry \cite{Banavar07,Banavar09}. And, as is well known, a garden hose is often wound into a helical geometry in a hardware store. Quite remarkably, a tightly wound tube has roughly the same pitch to radius ratio\cite{Maritan00} as an $\alpha$-helix in a protein and two tightly wound tubes in a double helix have the same geometry\cite{Stasiak00} as a DNA double-helix. We further break the resulting cylindrical symmetry by having side spheres of diameter $\sigma_{sc}$ sticking out on each main sphere in the negative normal direction. Our model is achiral, for simplicity. A chiral version of the model would have the side spheres in the plane perpendicular to the tangent and sticking out at a non-zero angle to the negative normal of the chain. The side spheres not only break the cylindrical symmetry of the overlapping spheres but also provide additional steric hindrance. The need to have the correct symmetry of a chain molecule has often been overlooked in the vast polymer science literature. 

The main chain monomers are subject to a generic short-range all/nothing attractive interaction. Any pair of monomers lying within a threshold distance, $R_c$, of each other is assigned a favorable unit of energy. In order to determine the ground state, one seeks to find the conformation with the minimum energy or, equivalently, the largest number of contacts within the threshold distance. There is one energy scale and four length scales in the model. The magnitude of the energy scale plays no role in determining the ground state phase diagram. Without loss of generality, we set the diameter of the main chain sphere to be 1 and measure the three other length scales in units of the main chain sphere diameter. The model that we study is entirely standard except for the somewhat unusual self-avoidance geometrical conditions, dictated by symmetry considerations. These alone lead to a rich and surprising phase diagram.

\section{Materials and methods}
\label{sec:materials}
In this Section we shall provide the details of the used formalism and the simulation protocol that has been used in the simulations.

\subsection{The model}
\begin{figure}[htpb]
  \centering
  \captionsetup{justification=raggedright,width=\linewidth}
  \begin{subfigure}{8cm}
     \includegraphics[width=.8\linewidth]{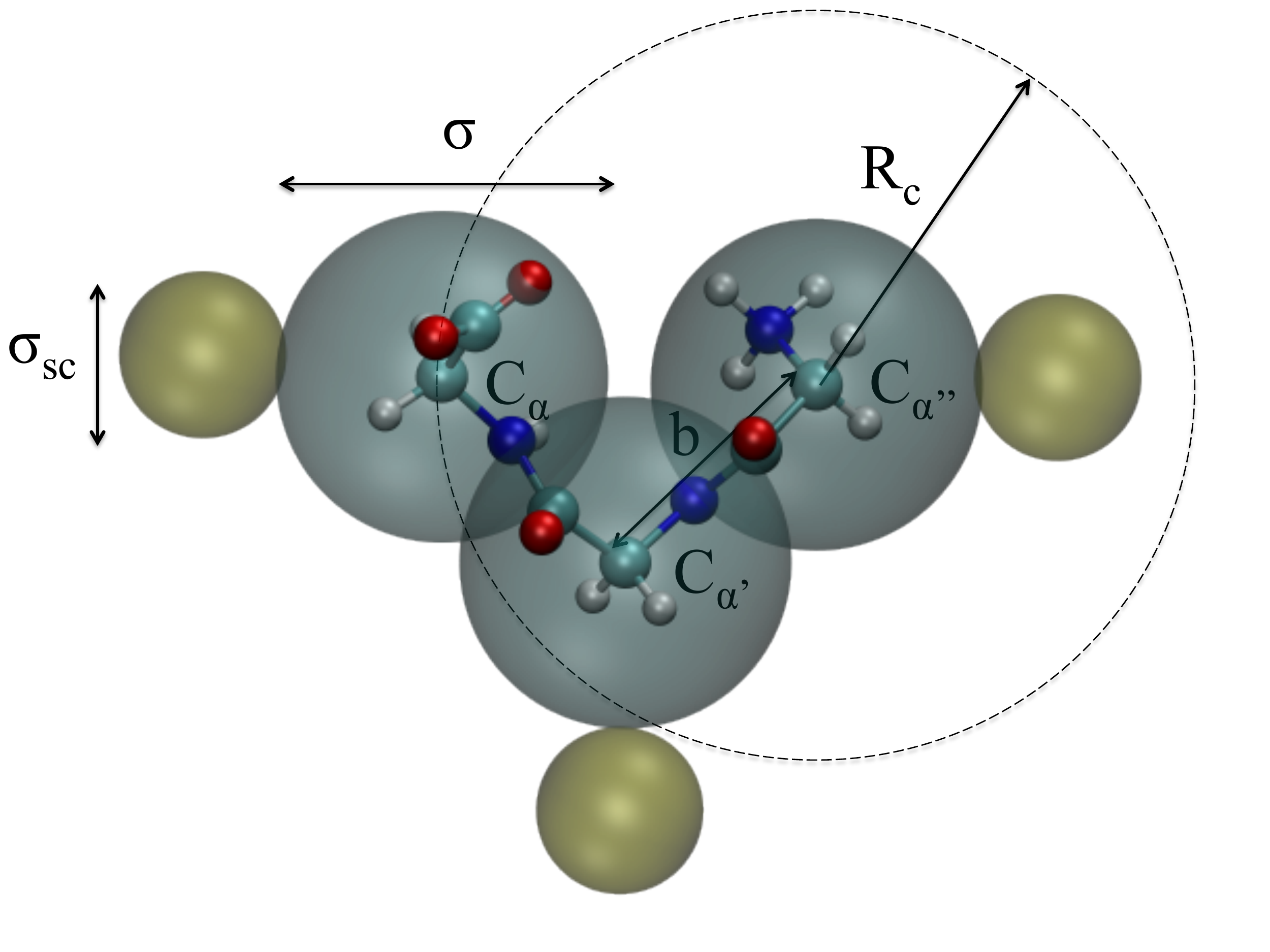}
    \caption{}\label{fig:fig1a}
  \end{subfigure}
  \begin{subfigure}{8cm}
     \includegraphics[width=.8\linewidth]{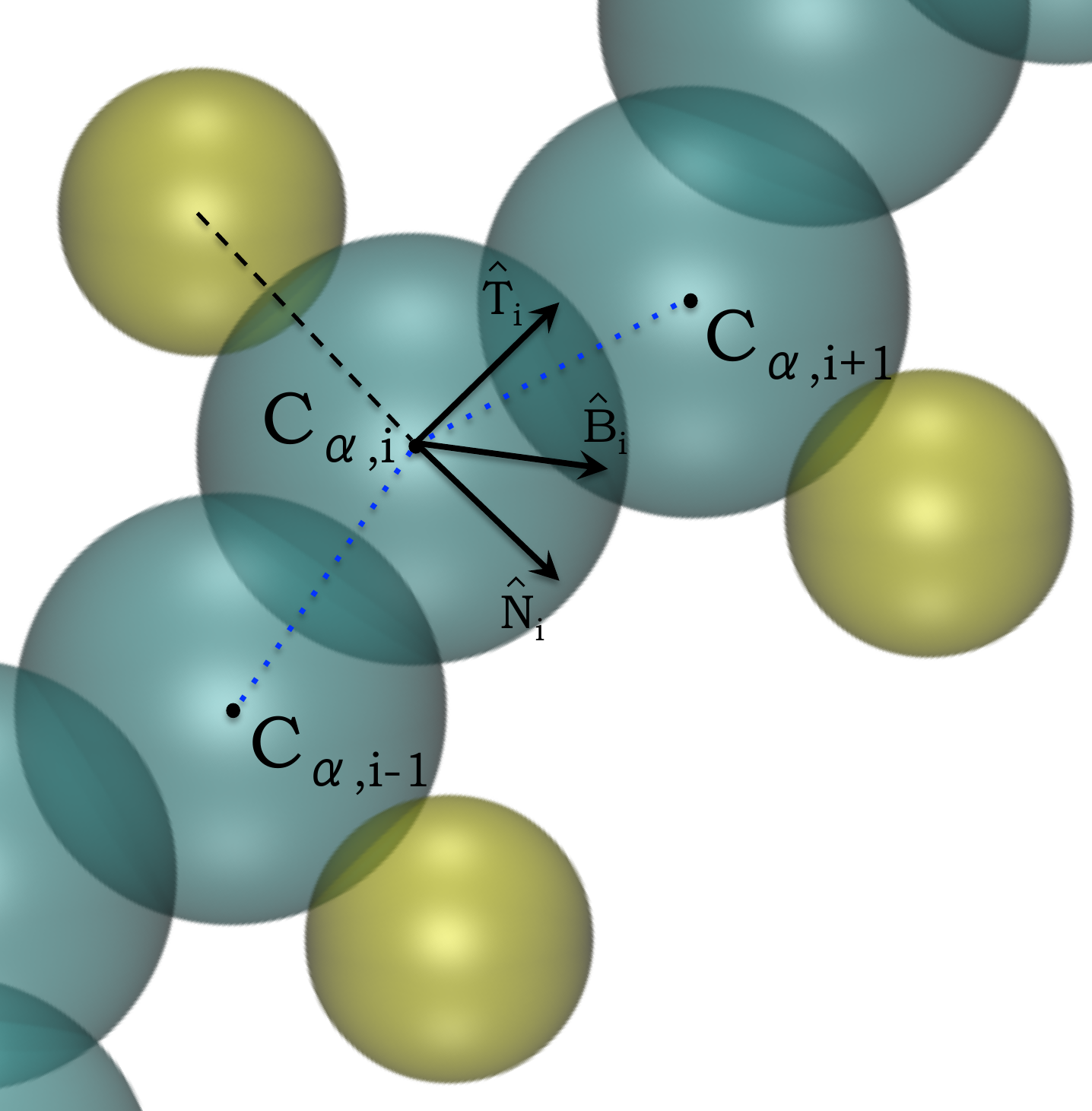}
    \caption{}\label{fig:fig1b}
  \end{subfigure}
  \caption{(a) Chain model. Each main chain sphere (cyan) has diameter $\sigma$. The side sphere (yellow) is in the negative normal direction and has a diameter $\sigma_{sc}$. The distance between successive main chain spheres is $b\le \sigma$ -- consecutive spheres can, in general, partially overlap. Non-consecutive main chain spheres experience a short range attractive constant potential if their separation is within the range of the attraction $R_c$. The atoms of glycine amino acids are shown as a guide. (b) Side sphere positions in Frenet frame.
  \label{fig:fig1}}
\end{figure}

Figure \ref{fig:fig1a} gives a sketch of our model. It consists of $N$ tethered spheres located at positions $\{ \mathbf{r}_1,\ldots,\mathbf{r}_N\}$, each having diameter $\sigma$ and a consecutive sphere-sphere distance along the chain of $b\le \sigma$. The case $b=\sigma$ corresponds to neighboring spheres along the chain just touching each other. Here consecutive main-chain spheres are in general allowed to interpenetrate to respect the correct (cylindrical)) symmetry imposed by the local tangent to the curve and in this case $b/\sigma <1$. Non-consecutive spheres have hard core interactions, as well as an attractive square-well interaction of range $R_c$ capturing non-directional generic hydrophobic interactions. Side spheres of diameter $\sigma_{sc}$ are added to each of the $N-2$ internal spheres along the negative normal direction, $-\widehat{\mathbf{N}}_i$, in the Frenet frame (see Fig \ref{fig:fig1b}) to capture the steric hindrance of the side chains and to break the cylindrical symmetry in an achiral manner. 
\subsection{The Frenet formalism}
The Frenet frame is defined by a tangent vector \cite{Kamien02}
\begin{eqnarray}
  \label{eq:eq1}
  \widehat{\mathbf{T}}_i&=&\frac{{\bm \Delta}_{i}+{\bm \Delta}_{i+1}}{\left \vert {\bm \Delta}_{i}+{\bm \Delta}_{i+1} \right \vert}
\end{eqnarray}
where ${\bm \Delta}_i=\mathbf{r}_{i}-\mathbf{r}_{i-1}$ so that $\vert{\bm \Delta}\vert=b$. For each of the non-terminal backbone spheres, one defines a normal vector 
\begin{eqnarray}
  \label{eq:eq2}
\widehat{\mathbf{N}}_i & = & \frac{{\bm \Delta}_{i+1}-{\bm{\Delta}_{i}}}
{\vert {\bm \Delta}_{i+1}-{\bm \Delta}_{i} \vert }
\end{eqnarray} 
where $i=2,\ldots,N-1$. The corresponding binormal vector is then given by
\begin{eqnarray}
  \label{eq:eq3}
\widehat{\mathbf{B}}_i &=& \widehat{\mathbf{T}}_i \times \widehat{\mathbf{N}}_i .
\end{eqnarray}
These equations are discrete versions of the continuum Frenet frame frequently used in polymer theory \cite{Kamien02}

Apart from their excluded volume, the side spheres do not interact either with the
backbone spheres or with each other. The side spheres merely have only steric interactions with main chain spheres and with each other.

The position of the side sphere is defined by
\begin{eqnarray}
  \label{eq:eq4}
\mathbf{r}^{(sc)}_i &=& \mathbf{r}_i - \widehat{\mathbf{N}}_i \frac{\left(\sigma + \sigma_{sc}\right)}{2} ;
\end{eqnarray}
the side chain sphere and the backbone sphere are tangent to each other. Amino acids in real proteins have an average tilt of roughly $41^{\degree}$ degrees with respect to the negative normal direction
and have different sizes. Our model does not have any chirality built into it.

\subsection{Simulations protocol}
We have studied the zero-temperature phase diagram using microcanonical Wang-Landau \cite{Wang01} and parallel tempering (also known as replica exchange) canonical Monte Carlo simulations \cite{Frenkel01}, always with consistent results.  In the Wang-Landau simulations, we sample polymer conformations according to the micro-canonical distribution by generating a sequence of chain conformations $A \to B$, and accepting the new configuration $B$ with the micro-canonical acceptance probability
\begin{eqnarray}
\label{eq:WL}
P_{acc}(A \rightarrow B)&=&\min{ \left(1,\frac{w_B \, g(E_A)}{w_A \, g(E_B)} \right)},
\end{eqnarray}
\noindent where $w_A$ and $w_B$ are weight factors ensuring the microscopic reversibility of the moves. 
The set of MC moves, that are accepted or rejected according to the probability (\ref{eq:WL}) includes both local-type moves, such as single-sphere crankshaft, reptation and end-point, as well as non-local-type moves, such as pivot, bond-bridging and back-bite moves, randomly sampled so that on average $N$ spheres (or a multiple of it) are moved to complete a MC step.

The density of states $g(E)$ is then constructed iteratively by filling suitable energy histograms and controlling their flatness. However, in order to compute
the ground state energy, the lowest energy was consecutively selected using the acceptance probability (\ref{eq:WL}) with a bias toward less populated energy states. This corresponds to the usual preliminary calculation carried out without a low-energy cut-off in the usual Wang-Landau scheme. In the full Wang-Landau calculation, we typically assume convergence after 30 levels of iterations, corresponding to a multiplicative factor value of $f=10^{-9}$.
For the ground state calculation, each run is composed of at least $10^9$ Monte Carlo moves per chain.

In the parallel tempering simulations, we carried out individual canonical Monte Carlo simulations within a Metropolis scheme at fixed temperature, with periodic swapping between neighboring temperatures within a prescribed annealing schedule. The acceptance probability for the exchange between two
replicas $\Gamma_i$ and $\Gamma_j$  at temperatures $T_i$ and $T_j$ respectively, is given by the acceptance probability
\begin{eqnarray}
  \label{eq:RE}
P_\mathrm{swap} &=& \min \left(1,\exp \left[ \left(\frac{1}{k_B T_i} - 
\frac{1}{k_B T_j}\right)(E_i - E_j)\right] \right).
\end{eqnarray}
12 replicas were used with the replica temperatures expanding from those of the swollen phase to those of the compact phase. The total number of steps ranged from $10^8$ to $10^9$ sweeps per replica, depending on the system size, one step corresponding to $N$ attempted MC moves, where $N$ is the number of beads. A replica exchange is attempted every 10 Monte Carlo steps. The move sets include pivot, crankshaft, and reptation moves with probabilities $0.2$, $0.7$ and $0.1$, respectively.

The results obtained are robust and independent of the technique employed. 
\section{Results}
\label{sec:results}

\begin{figure}[htpb]
  \centering
  \captionsetup{justification=raggedright,width=\linewidth}
  \begin{subfigure}{8cm}
     \includegraphics[width=.8\linewidth]{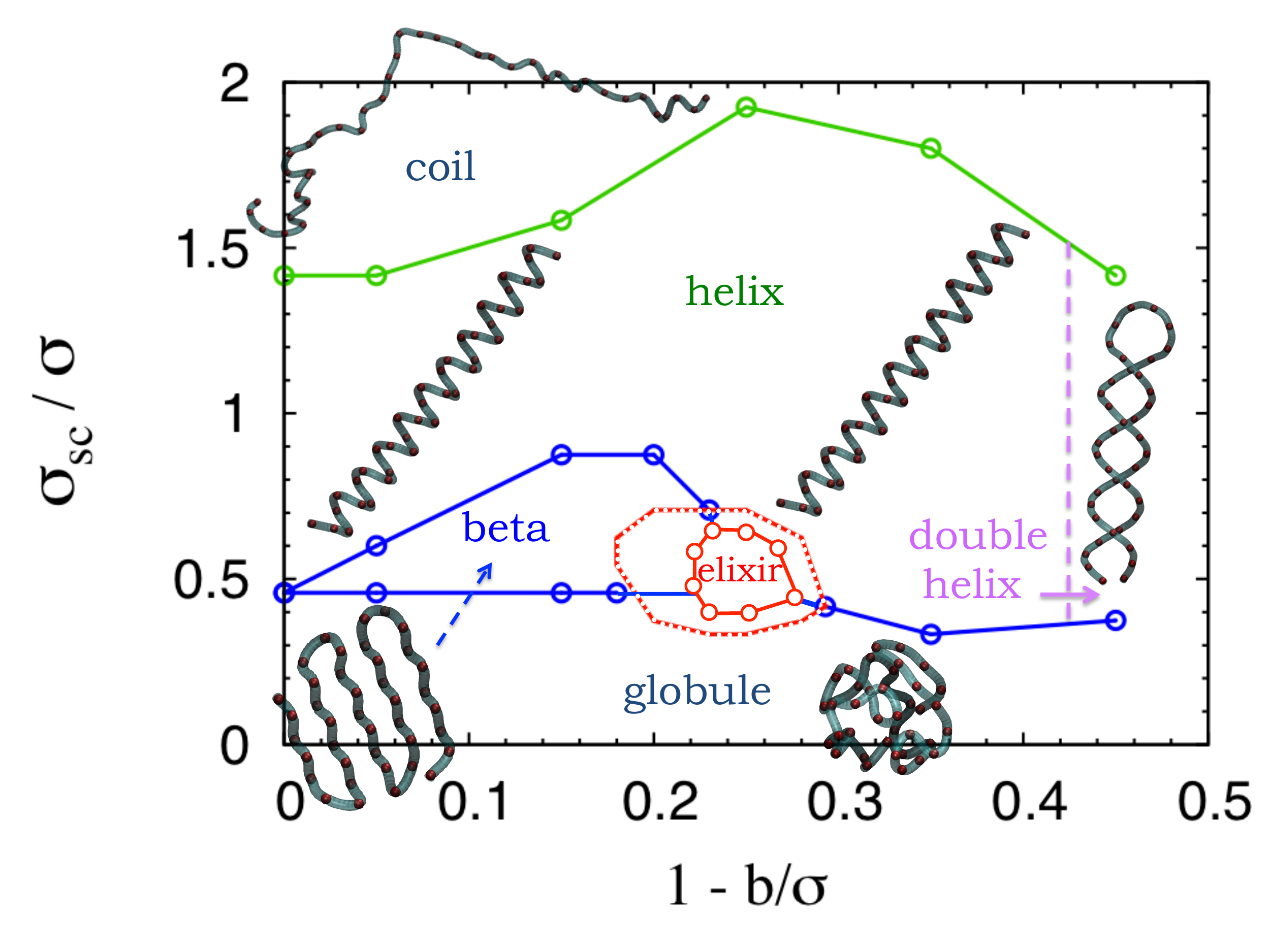}
    \caption{}\label{fig:fig2a}
  \end{subfigure}
  \begin{subfigure}{8cm}
     \includegraphics[width=.8\linewidth]{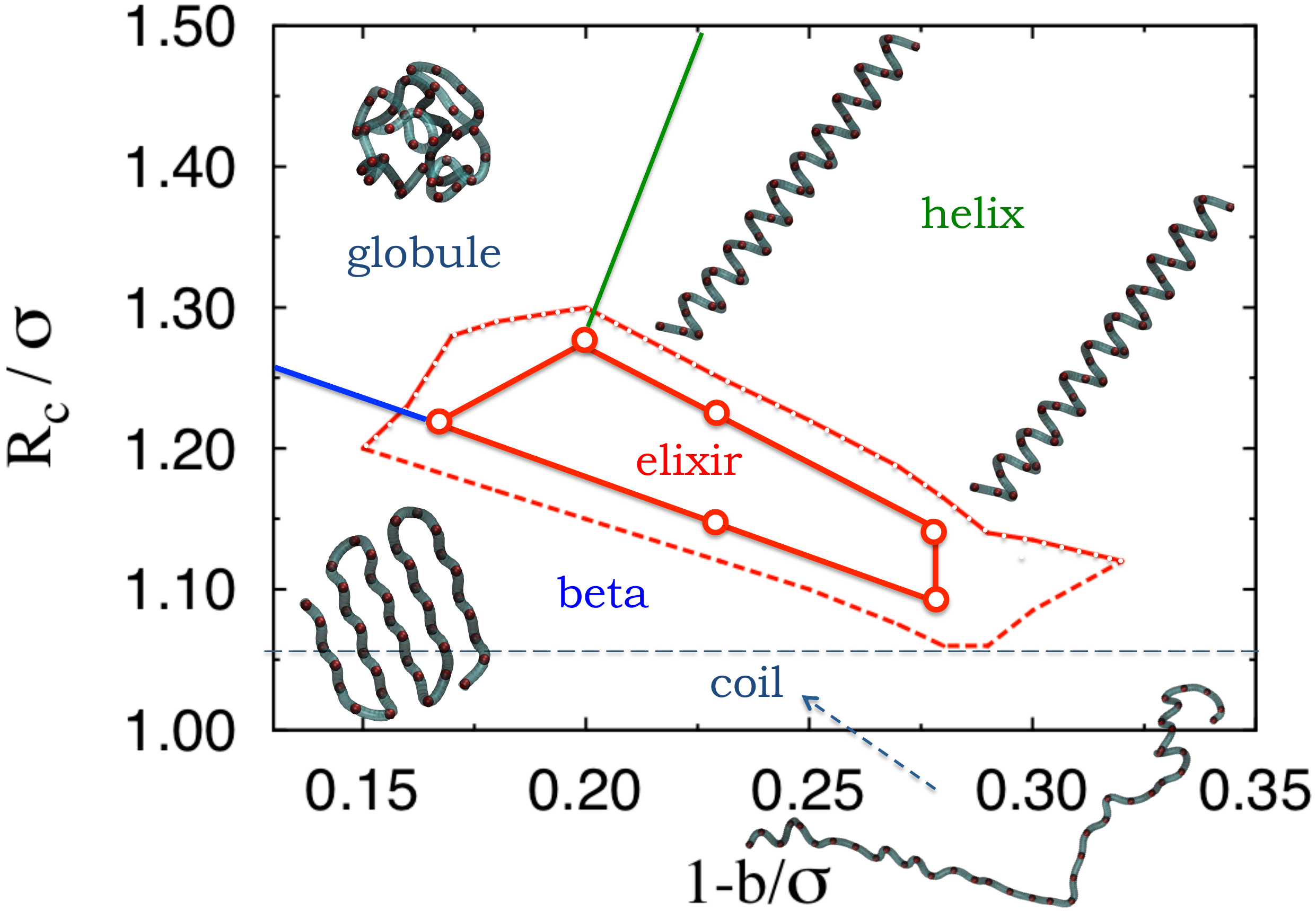}
    \caption{}\label{fig:fig2b}
  \end{subfigure}\\
  \begin{subfigure}{8cm}
     \includegraphics[width=.8\linewidth]{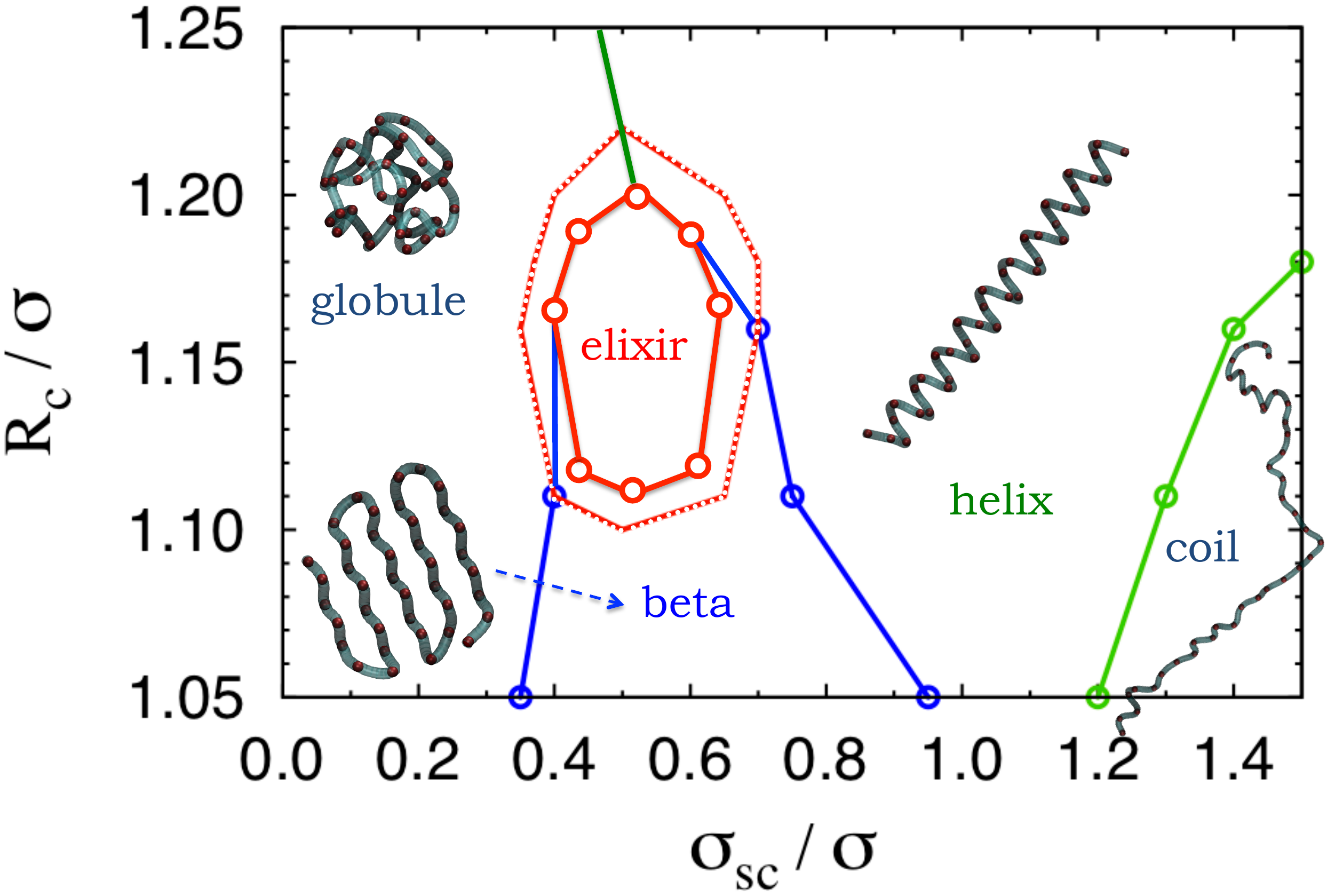}
    \caption{}\label{fig:fig2c}
  \end{subfigure}
  \begin{subfigure}{9cm}
    \includegraphics[width=.8\linewidth]{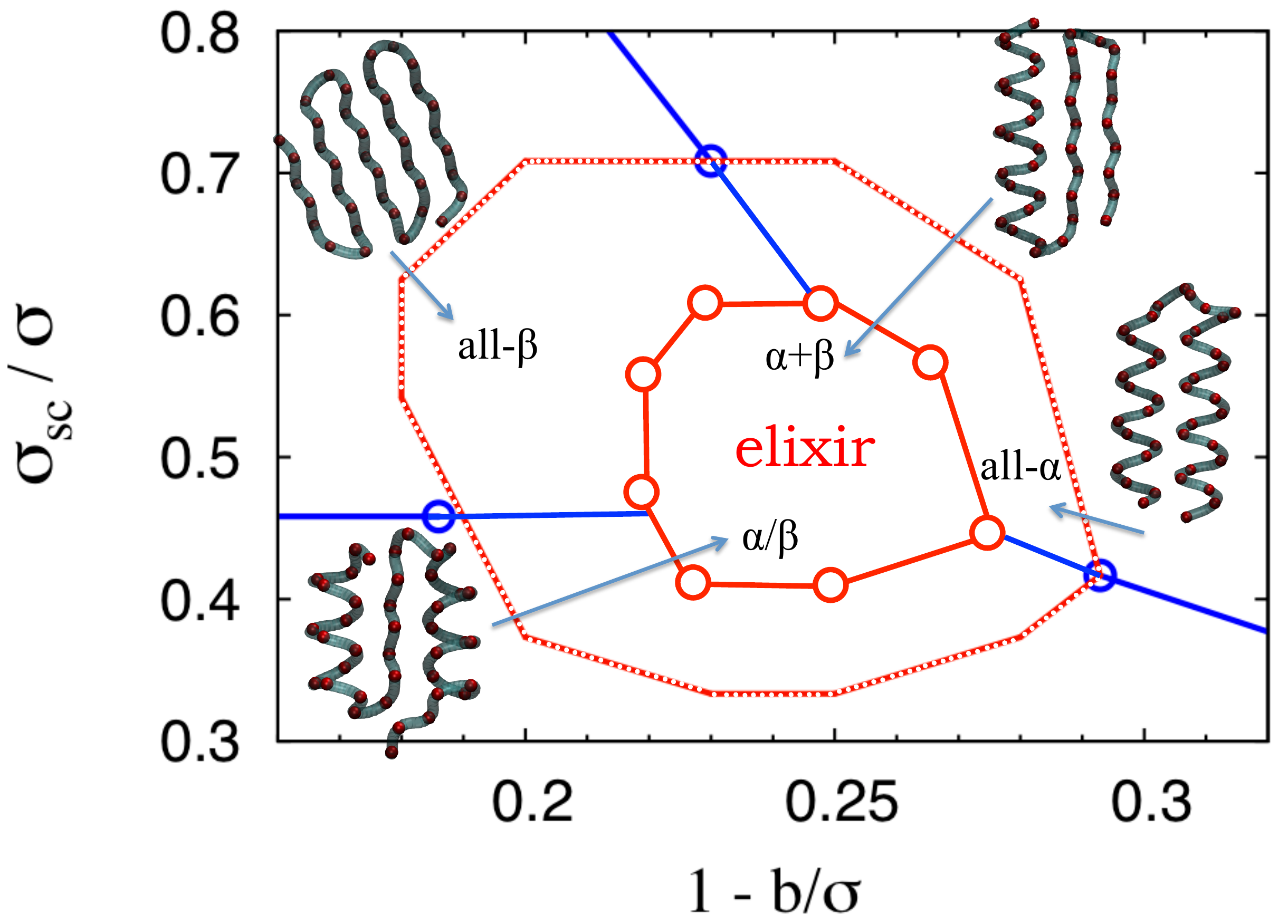}
    \caption{}\label{fig:fig2d}
  \end{subfigure}  
  \caption{Phase diagram of chain model. The main chain sphere diameter $\sigma$ sets the characteristic length scale of the model and is taken to be $5 \mathrm{\AA}$. The three dimensionless variables studied are: 1) the ratio of the diameters of the side-chain ($\sigma_{sc}$) and main chain ($\sigma$) spheres; 2) the ratio of the range of the attractive interaction ($R_c$) to the diameter of the main chain sphere ($\sigma$); and 3) the degree of overlap of consecutive main chain spheres measured by the difference between the diameter of the main chain sphere ($\sigma$) and the distance between the centers of consecutive main chain spheres along the chain ($b$) normalized by the diameter of the main chain spheres ($\sigma$). Panels \ref{fig:fig2a}-\ref{fig:fig2c} display the ground state phase diagrams in the three corresponding planes for a chain of length 40. Panel \ref{fig:fig2a} Side-chain size $\sigma_{sc}/\sigma$-overlap $1-b/\sigma$ plane. The range of interaction has been fixed at $R_c/\sigma=1.16$, corresponding to $R_c \approx 6 \mathrm{\AA}$. Panel \ref{fig:fig2b} Range of attraction $R_c/\sigma$- overlap  $1-b/\sigma$ plane. The side chain sphere size has been fixed to $\sigma_{sc}/\sigma=0.5$ corresponding to $\sigma_{sc} \approx 2.5 \mathrm{\AA}$. Panel  \ref{fig:fig2c} Range of attraction $R_c/\sigma$-side-chain sphere size $\sigma_{sc}/\sigma$ plane. The overlap value has been fixed at $1-b/\sigma=0.25$ corresponding to $b \approx 3.8 \mathrm{\AA}$. In all cases, the central enclosed phase, denoted as the \textit{elixir phase}, include nearly \textit{degenerate} (in energy) conformations comprised of combinations of $\alpha$ helices and $\beta$ strands. The larger dotted enclosed region, contain, in addition, conformations having either $\alpha$ helices or $\beta$ strands with geometries matching those appearing in real proteins. These also include all-$\beta$ and all-$\alpha$ conformations that are not part of the elixir phase (Panel \ref{fig:fig2d}).
  \label{fig:fig2}}
\end{figure}

The self-avoidance is unusual (and motivated by protein chemistry) compared to previous studies because of the lack of symmetry, which is a feature that has been often overlooked. The phase diagram is a result of the competition between the large number of self- avoiding conformations (entropy) and the few self-avoiding compact conformations, which avail of the attractive interactions and is a transition driven by temperature. The elixir phase persists over a range of temperatures and is a low temperature phase. Our model is for a single chain. Unlike, a system of hard spheres where packing fraction is a key parameter, the density does not play a role here.
The microcanonical Wang-Landau calculations allow one to measure the density of states and hence the free energy from which all thermodynamics can be obtained. The same method also provides a direct way to access the “ ground state” of the system. In parallel tempering, we gradually lower the temperature and monitor the energy. Below the folding transition temperature, this is taken as the “ground state” energy and matches the value obtained via the Wang-Landau calculation. Taken together, these methods confirm that the elixir phase is a low temperature phase.
The elixir phase is stable over a range of low temperatures. When the system is cooled below the folding transition temperature, it achieves the analog of the native state in real proteins. We denote this as a ground state and Figure \ref{fig:fig2} displays the corresponding phase diagram.
In addition to conventional polymer phases, the phase diagram has regions with a unique ground state including a single $\beta$-sheet, a single helix, and two helices wrapped around each other. Nestled between these phases lies a particularly interesting phase, that we have denoted as the \textit{elixir phase}, whose ground states are assembled structures of helices and strands. The elixir phase is degenerate with the assembled structures having nearly the same energy (see below). Intriguingly, the geometries of the building block strands and helices in the elixir phase, as well as those in a larger surrounding region identified by dotted lines in Figure \ref{fig:fig2d}, are statistically the same as those of strands and helices in protein native state structures. Each ground state in the phase diagram was assigned to a specific phase on the basis of suitable order parameters \cite{Skrbic16b} that includes the twist (for the helices) and the average triple scalar product of normal Frenet unit vectors (for the sheets). An additional fingerprint of the secondary structures stems from their characteristic representations in the contact maps.
The figures label the phases and also depict the types of structures supported in each of the phases. The entire region between the coil and the globule phases shrinks to a single point for a conventional chain model with no overlap between consecutive main chain spheres and with no side spheres. This underscores the key roles of breaking, first, the spherical symmetry (capturing the correct local cylindrical symmetry associated with any chain molecule through the overlap) and then the cylindrical symmetry (through the side spheres). The phases labeled as helix and beta have ground states of a single helix and a single beta sheet, respectively.

\begin{figure}[htpb]
  \centering
  \captionsetup{justification=raggedright,width=\linewidth}
  \begin{subfigure}{8cm}
     \includegraphics[width=.8\linewidth]{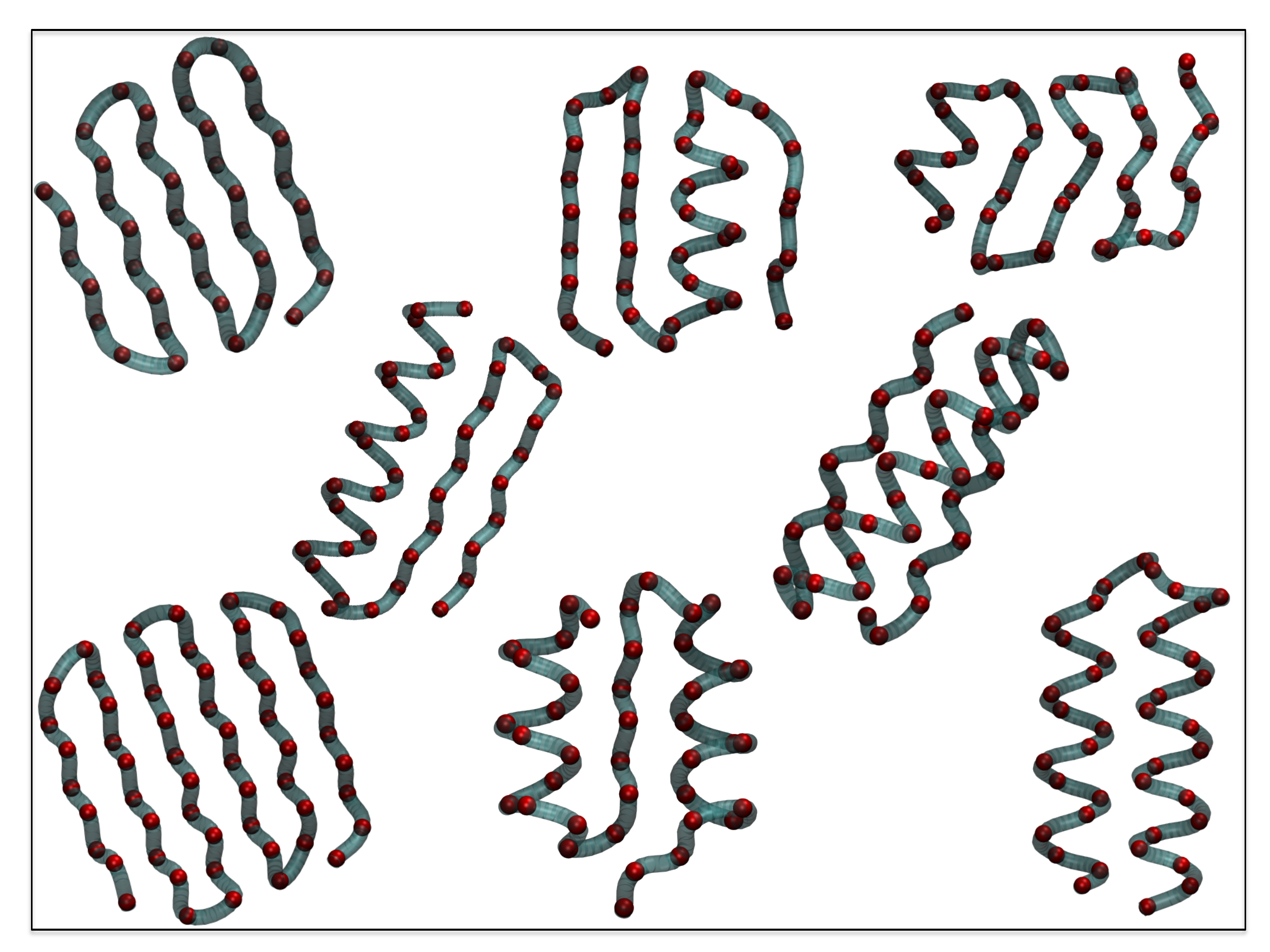}
    \caption{}\label{fig:fig3a}
  \end{subfigure}
  \begin{subfigure}{8cm}
     \includegraphics[width=.8\linewidth]{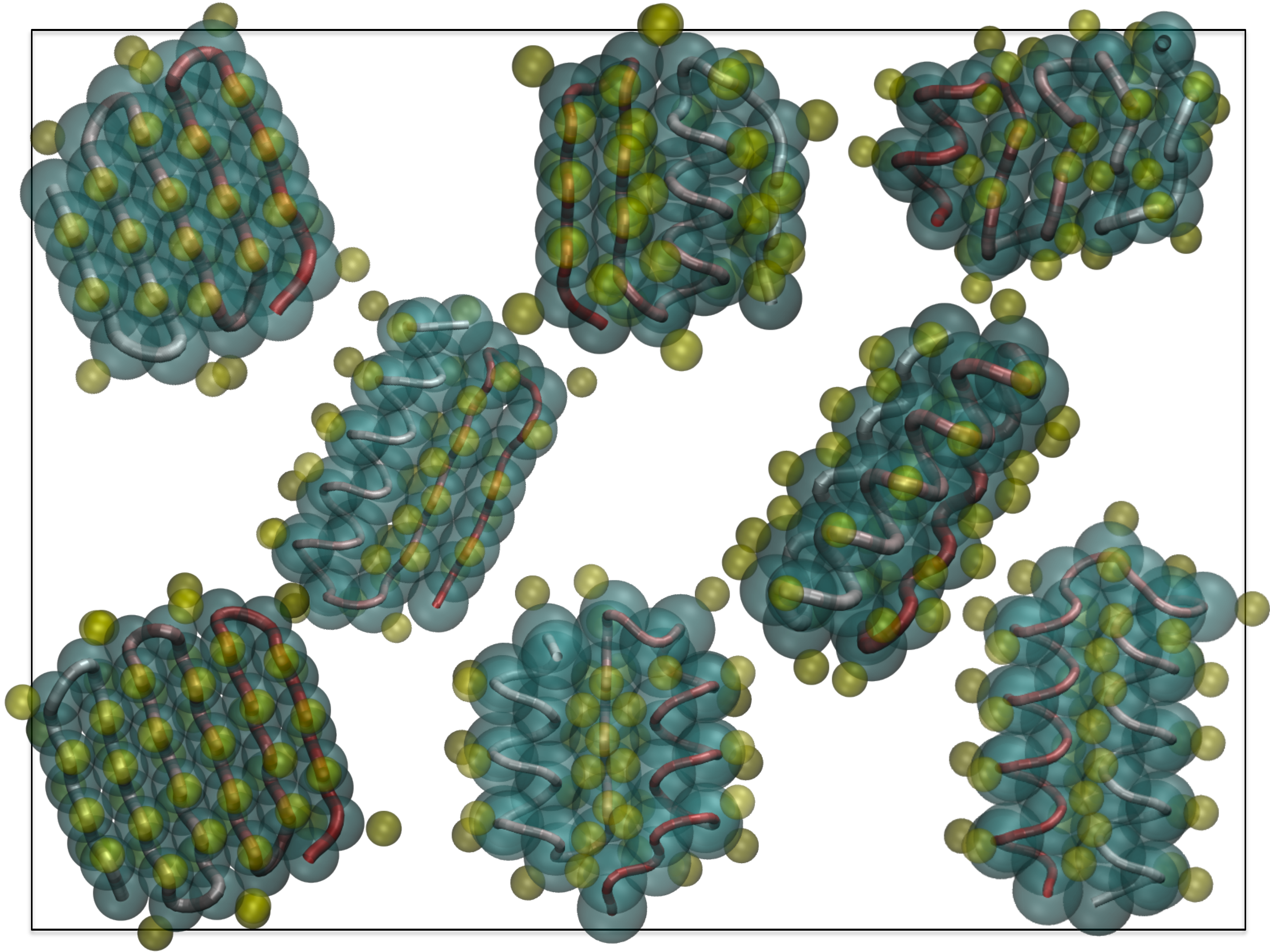}
    \caption{}\label{fig:fig3b}
  \end{subfigure}
  \caption{Panel \ref{fig:fig3a} is a gallery of ground state structures found in the elixir phase and in the extended region surronding it. All helices have radii $(2.4 \pm 0.1)\mathrm{\AA}$ and pitches $(5.5 \pm 0.5) \mathrm{\AA}$ matching those of protein helices within error bars. Likewise, all strands have distances between $C_{\alpha}^{i}$ and $C_{\alpha}^{i+2}$ $\approx 6 \mathrm{\AA}$, as in proteins. A tube-like representation of the chain has been employed for better visualization. Panel  \ref{fig:fig3b} is the same as in Panel \ref{fig:fig3a}, but with full display of the spheres. The ground state structures within the elixir phase and those shown in Panels \ref{fig:fig3a} and \ref{fig:fig3b} are low energy structures with approximately the same number of contacts.
  \label{fig:fig3}}
\end{figure}

The elixir phase is remarkable in several ways: it is stable for moderate length chains -- we find no significant change in the nature of the phase on varying the chain length between 20 and around 200; the nearly degenerate ground state structures are geometrical sculptures assembled from building blocks of helices and sheets allowing for geometry based lock-key interactions so ubiquitous in living systems; the assembled structures in the elixir phase have topologies similar to the observed folds in proteins --  $\alpha$+$\beta$ and $\alpha/\beta$ \cite{Levitt76,Richardson92} (see Figures \ref{fig:fig3a} and \ref{fig:fig3b} for a gallery of conformations); and the elixir phase is surrounded by other phases and thus the structures within it are poised in the vicinity of phase transitions albeit for small systems. 

\begin{figure}[htpb]
  \centering
  \captionsetup{justification=raggedright,width=\linewidth}
  \begin{subfigure}{8cm}
     \includegraphics[width=.8\linewidth]{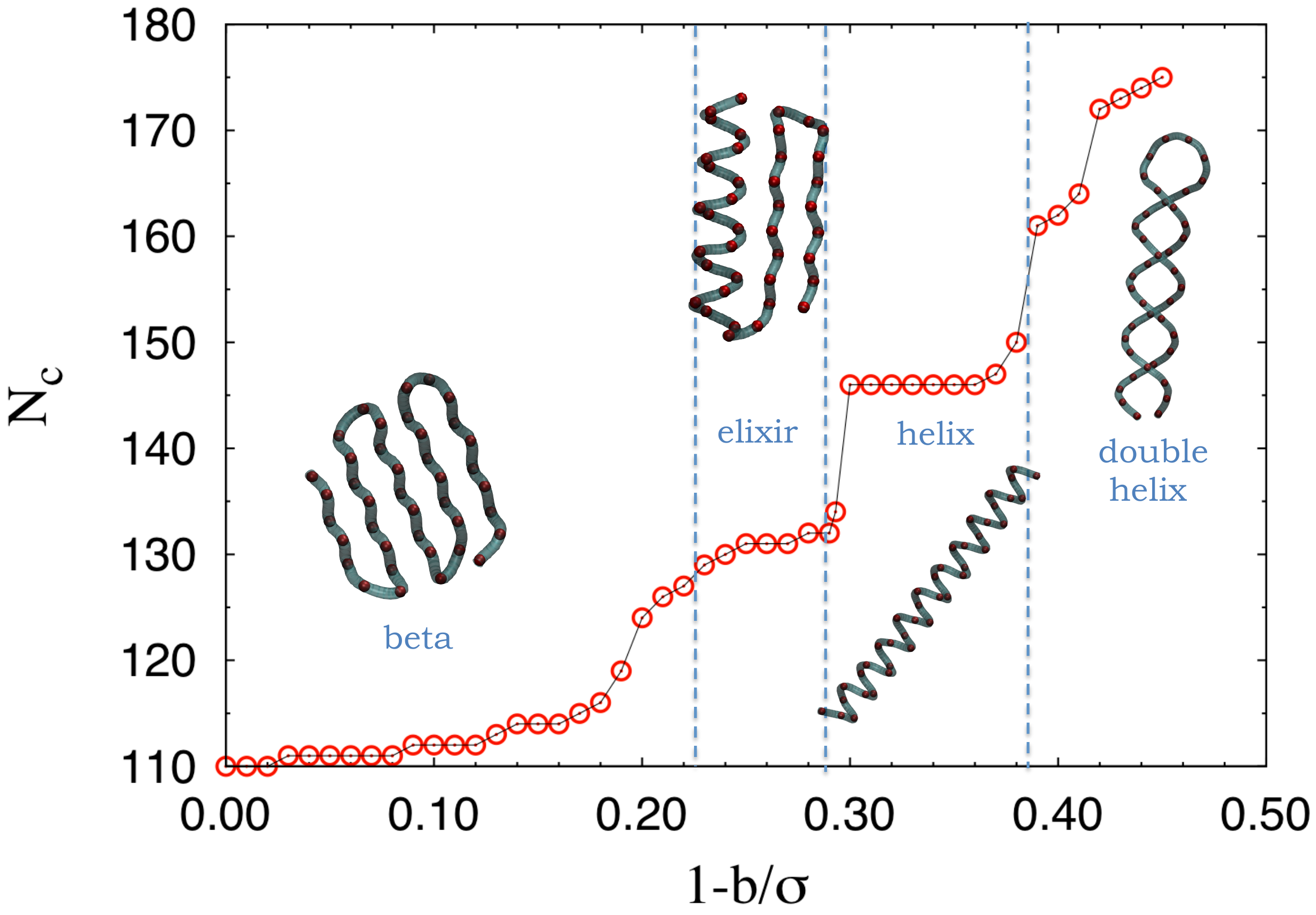}
    \caption{}\label{fig:fig4a}
  \end{subfigure}
  \begin{subfigure}{8cm}
     \includegraphics[width=.8\linewidth]{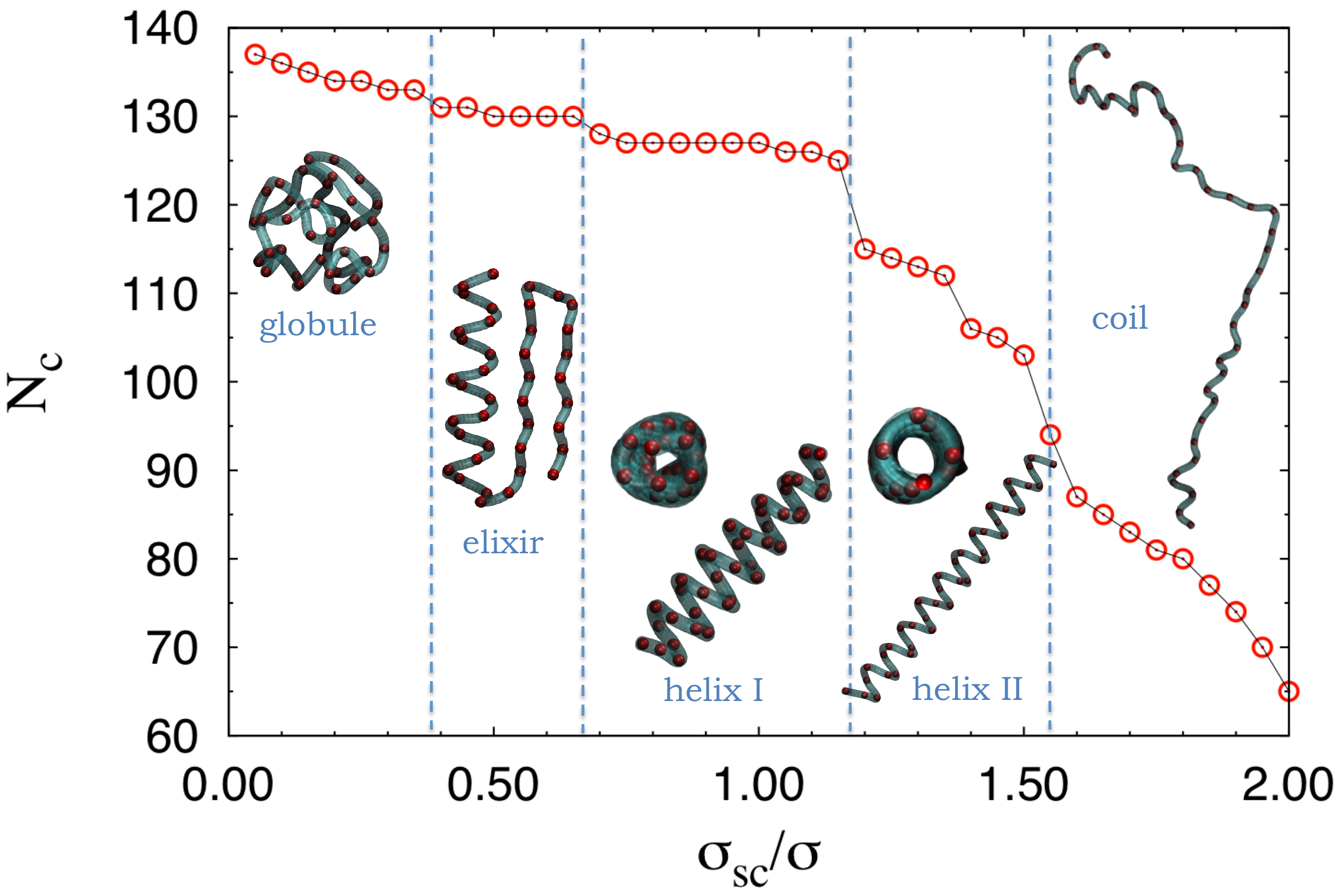}
    \caption{}\label{fig:fig4b}
  \end{subfigure}\\
  \begin{subfigure}{8cm}
     \includegraphics[width=.8\linewidth]{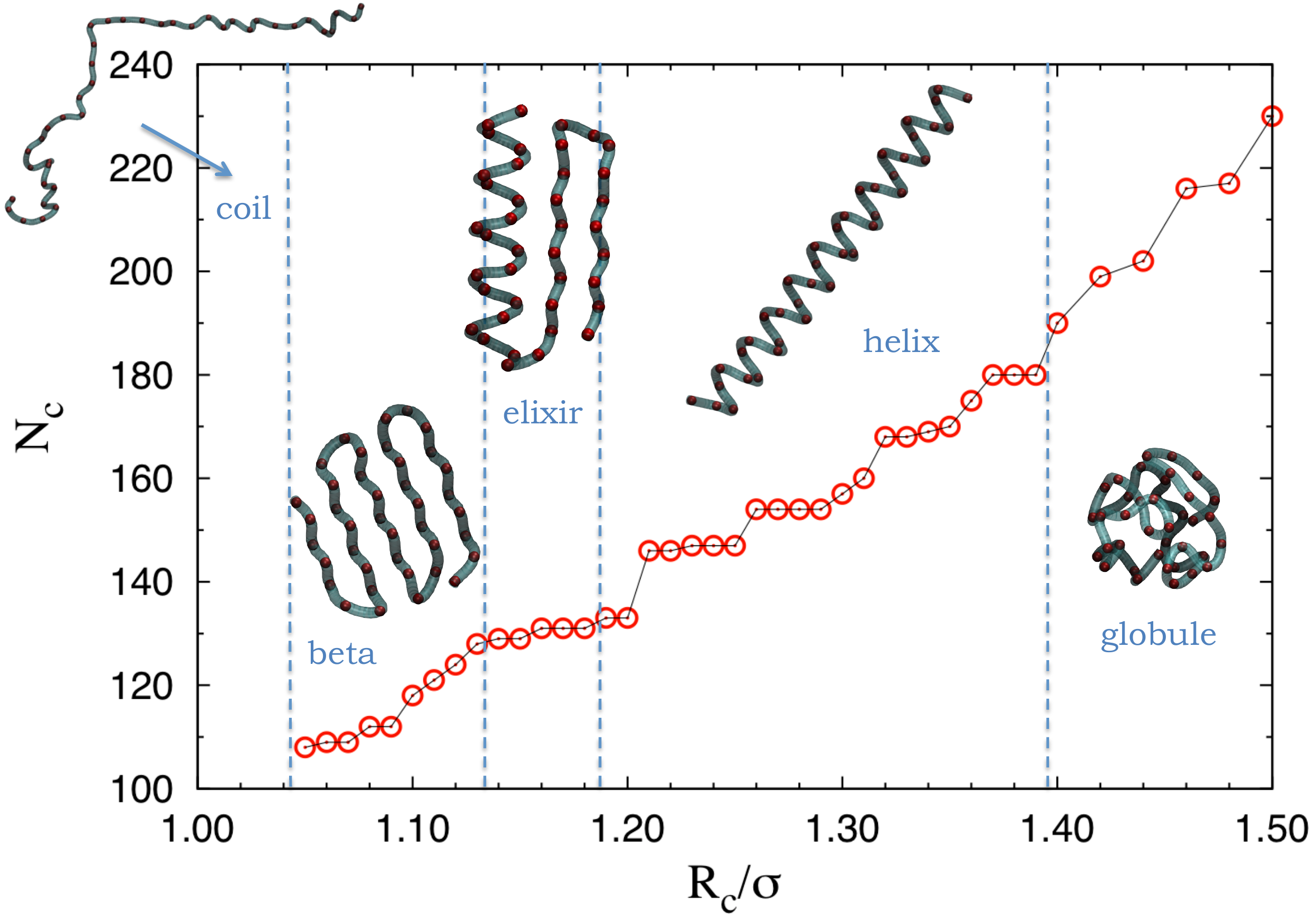}
    \caption{}\label{fig:fig4c}
  \end{subfigure}
  \begin{subfigure}{9cm}
    \includegraphics[width=.8\linewidth]{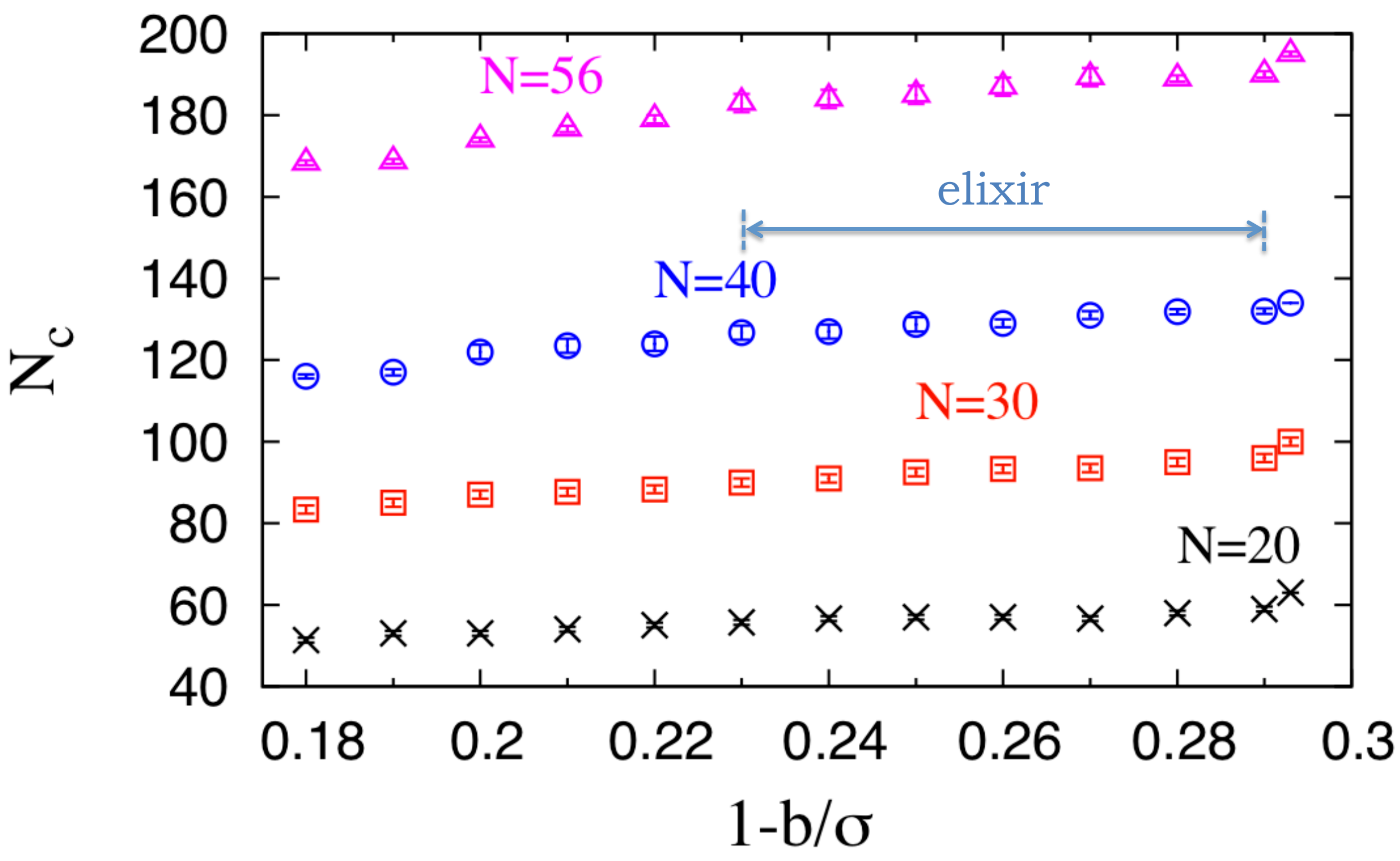}
    \caption{}\label{fig:fig4d}
  \end{subfigure}
  \begin{subfigure}{9cm}
    \includegraphics[width=.8\linewidth]{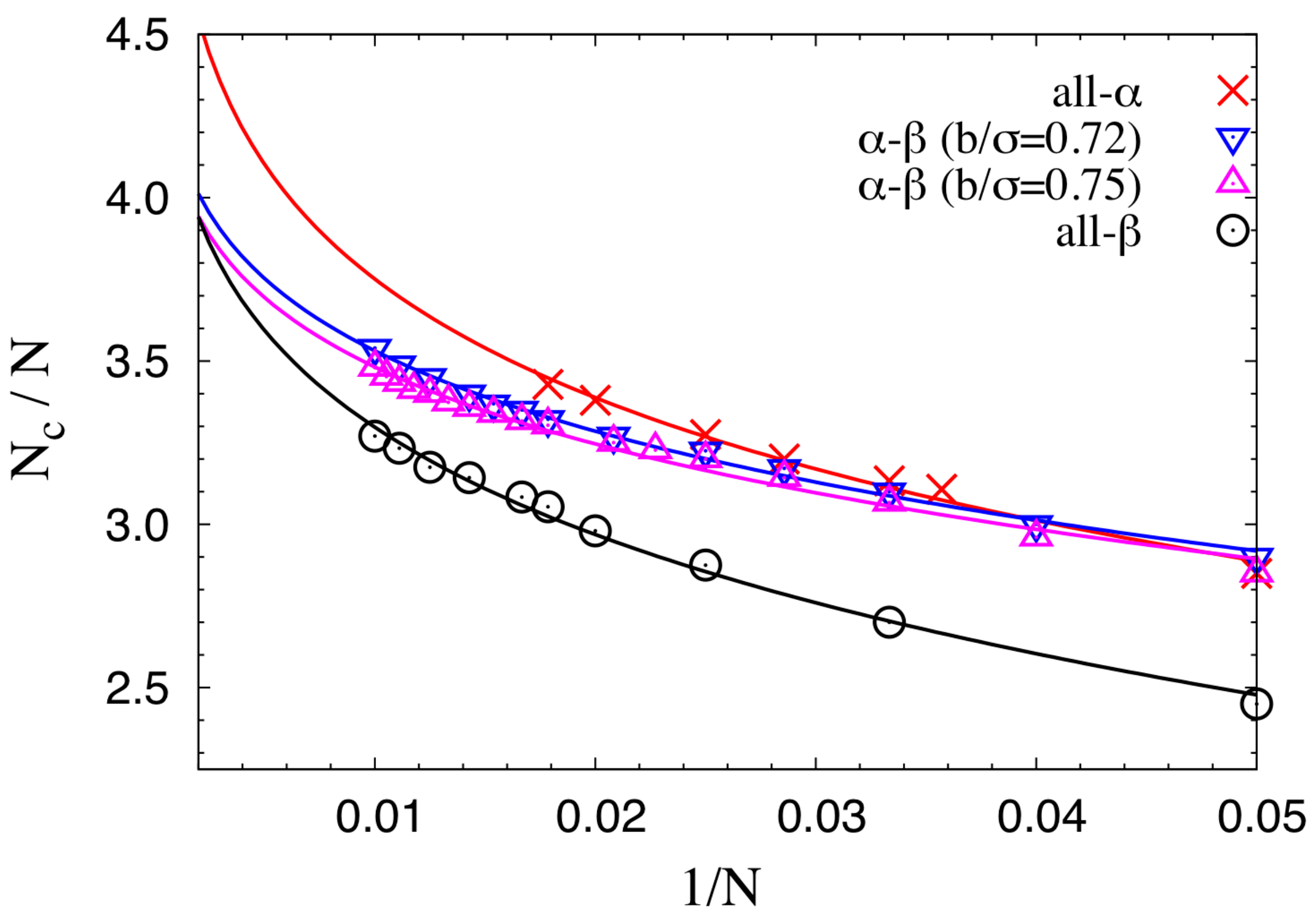}
    \caption{}\label{fig:fig4e}
  \end{subfigure} 
  \caption{Number of ground state contacts $N_c$ as a function of $1-b/\sigma$ (\ref{fig:fig4a}), $\sigma_{sc}/\sigma$ (\ref{fig:fig4b}) and $R_c/\sigma$ (\ref{fig:fig4c}) for a chain of length 40. The vertical lines indicate transitions to the phases housing conformations shown in the snapshots, and are consistent with the transitions of the phase diagrams in Figure \ref{fig:fig2}. The transition helixI$\to$helixII appearing in the central panel is a structural transition between two different type of helices, and, for simplicity, is not explicitly displayed in the phase diagram of Figure \ref{fig:fig2}. In all cases, the remaining two parameters have been set to the center of the elixir phase, as in Figure \ref{fig:fig2} of the main text: $1-b/\sigma=0.25$, $\sigma_{sc}/\sigma=0.5$, and $R_c/\sigma=1.16$. Figure (\ref{fig:fig4d}) shows the number of ground state contacts $N_c$ as a function of $1-b/\sigma$ for different chain lengths ($N=20,30,40,56$). Each point is the average over $5-10$ independent runs, with errors bars of the order of the size of each point.  For $N=40$ the range of the elixir phase is highlighted. Figure (\ref{fig:fig4e}) shows the number of ground state contacts per bead $N_c/N$ for large $N$, showing that it tends to a value of $\approx 4$. Note that the all$\beta$ conformations tend to extrapolate to the same value, whereas the all-$\alpha$ conformations approach a slightly higher value of $\approx 4.5$.
  \label{fig:fig4}}
\end{figure}

Figure \ref{fig:fig4} shows the total number of ground state contacts, $N_c$, as a function of the three characteristic length scales of the model.  The boundaries of the elixir phase in Figures \ref{fig:fig2a}-\ref{fig:fig2d} correspond to the vertical dashed lines reported in Figure \ref{fig:fig3}. For positive increasing $R_c/\sigma$, the number of contacts $N_c$ monotonically increases from very small values characteristic of the coil phase until a threshold value sufficiently large to form a $\beta$ phase as shown in Figure \ref{fig:fig4c}. The geometry of the strands, as measured by the angle subtended at each main chain sphere with its adjacent spheres along the chain, changes smoothly until it becomes similar to that of a protein strand at $R_c/\sigma \approx 1.1$. Upon increasing $R_c/\sigma$ further, one enters the degenerate elixir phase with combined $\alpha$-$\beta$ structures, where the number of contacts is essentially independent of $R_c/\sigma$ as indicated by the flatness of $N_c$ in this region. At $R_c/\sigma \approx 1.20$, an abrupt upswing of $N_c$ signals the end of the elixir phase and entry into a single helix phase.

A similar phenomenology is observed along the $\sigma_{sc}/\sigma$ axis, as shown in Figure \ref{fig:fig4b}. Here $N_c$ monotonically decreases from the globular phase until it becomes flat upon entering into the elixir phase. The discontinuities seen within the helix phase correspond to a structural transition between two different geometries of helix (that we denote as helixI and helixII). The behavior of $N_c$ as a function of $1-b/\sigma$, shown in Figure \ref{fig:fig4a}, has a single beta phase for $1-b/\sigma>0$ (with $\sigma_{sc}/\sigma=0.5$ and $R_c/\sigma=1.16$). Again, the strand geometry changes smoothly and matches that of protein strands around $1-b/\sigma \approx 0.18$, as one approaches the elixir phase. The initial assembled topology is an all-$\beta$ conformation (see Figure \ref{fig:fig2d}), and $N_c$ progressively increases until a combined $\alpha$-$\beta$ ground state conformational topology is achieved, where $N_c$ is again approximately flat. At $1-b/\sigma \approx 0.29$, a drastic conformational change to a single helix is observed, where $N_c$ becomes flat again in the range $0.29 \le 1-b/\sigma \le 0.39$ characterized by maximally packed helices with consecutive turns lying one on top of the other.


A characteristic feature of the elixir phase is its approximate degeneracy, with many different ground state conformations with $N_c \approx 130$ for chain length $N=40$ (see Figure \ref{fig:fig4d} for different chain lengths $N$ going beyond the boundaries of the elixir phase). Figure \ref{fig:fig4e} shows that the number of ground state contacts per main chain sphere, $N_c/N$, tends to a value of $\approx 4$ in the elixir phase, in the limit of large $N$. Having building blocks (helices and beta strands) with the correct interplay between the geometries and range of interactions \cite{Wang15}, distinct self-assembled ground state topologies have comparable energies. Thus one can construct a library of many different topological folds in the elixir phase, all having energies within a small interval of $\approx 10\%$ spread (see Figures \ref{fig:fig3a} and \ref{fig:fig3b} for a representative sample).

\begin{figure}[htpb]
  \centering
  \captionsetup{justification=raggedright,width=\linewidth}
     \includegraphics[width=.8\linewidth]{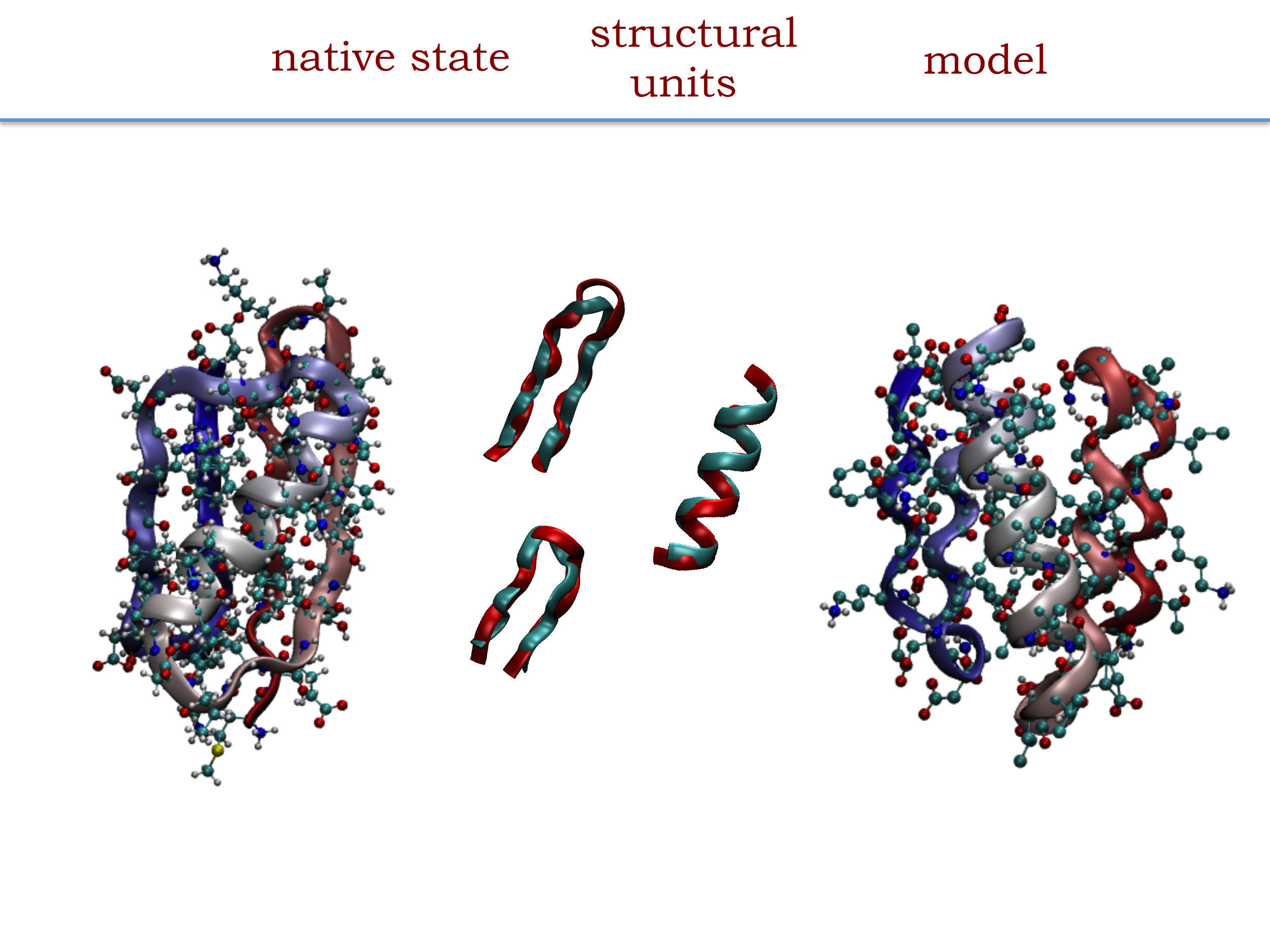}
  \caption{An example of comparison between ground state structures in the elixir phase and the native folds of proteins with $N=56$. Protein G (PDBid 3GB1), whose native state is formed by two $\beta$ antiparallel hairpins connected by a single $\alpha$ helix (left) and the topologically similar ground state in the elixir phase (right) with parameters: $\sigma = 5 \mathrm{\AA}$, $1-b/\sigma=0.25$, $\sigma_{sc}/\sigma=0.416$ and $R_c/\sigma = 1.16$. The root-mean-square deviation (RMSD) between the native state of the protein and the ground state of the model is $\approx 7 \mathrm{\AA}$. The central panel shows the fidelity of the overlap of the building blocks of the protein structures ($\alpha$ helices and $\beta$ hairpins) to those in the elixir phase (RMSD) $\le 2.0 \mathrm{\AA}$. Structural units of real proteins are shown in red, those from the model are in cyan. 
  \label{fig:fig5}}
\end{figure}

Figure \ref{fig:fig5} illustrates the remarkable similarity between the native fold of a protein and the corresponding ground state found in the elixir phase. Here, we have taken Protein G as representative example, but this result is valid for essentially \textit{any} topology. The left panel of Figure \ref{fig:fig5} represent its real native state. The right panel of Figure \ref{fig:fig5} shows the ground state conformation in the elixir phase, where the specific sequence pertaining to protein G has been inserted by using the PULCHRA tool \cite{Rotkiewicz08}. The central panel shows the overlap of the single units, both the $\alpha$ helix and the $\beta$ sheets, illustrating how they have the correct geometries matching those of real proteins.

\begin{figure}[htpb]
  \centering
  \captionsetup{justification=raggedright,width=\linewidth}
  \begin{subfigure}{8cm}
     \includegraphics[width=.8\linewidth]{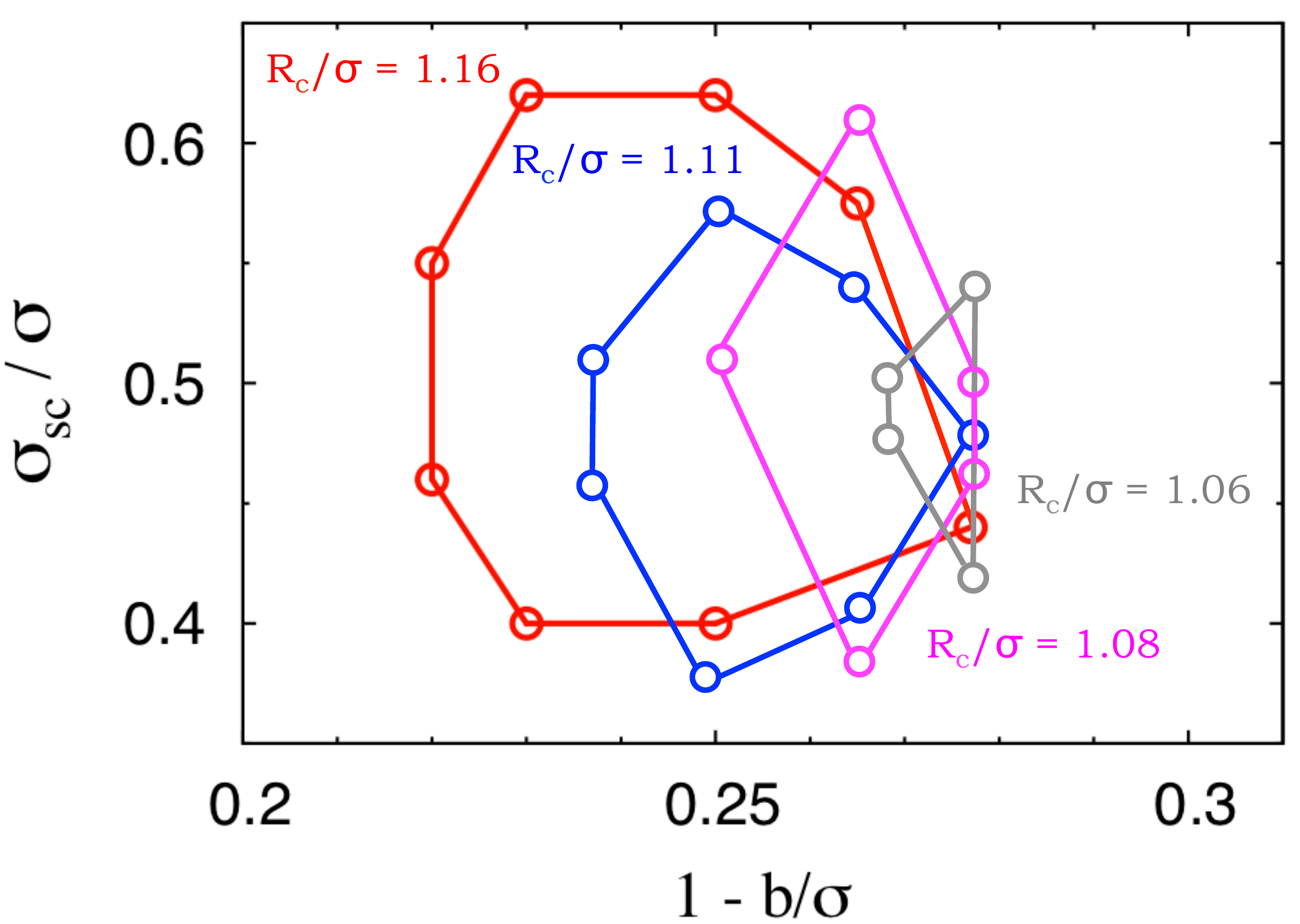}
    \caption{}\label{fig:fig6a}
  \end{subfigure}
  \begin{subfigure}{8cm}
     \includegraphics[width=.8\linewidth]{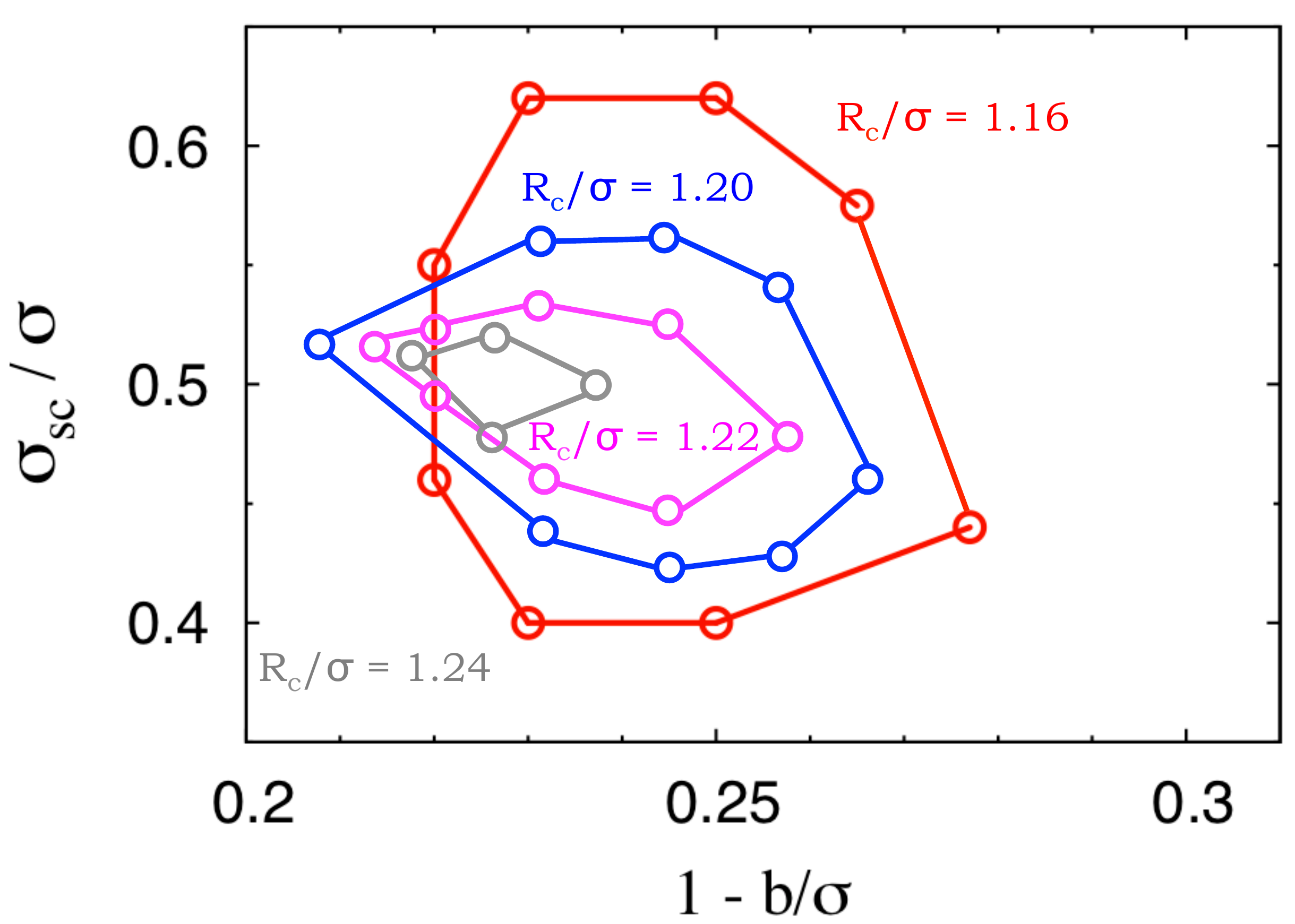}
    \caption{}\label{fig:fig6b}
  \end{subfigure}
  \caption{(a) Schematic view of the gradual shrinking of the elixir phase for a chain of length $40$ upon decrease of the attraction range, $R_c$. The red line corresponds to the boundary of the  elixir phase of Figure \ref{fig:fig2d}. Note the tendency to form a vertical line rather than a point in the limit $R_c/\sigma \to 1$ .(b) Same upon increasing $R_c$. Here, unlike in the opposite limit, the elixir phase tends to shrink to a point and eventually disappear.
  \label{fig:fig6}}
\end{figure}

We stress the importance of selecting the correct value of $R_c/\sigma \approx 1.16$ (corresponding to $R_c \approx 6 \mathrm{\AA}$ for $\sigma \approx 5 \mathrm{\AA}$) in order to observe the elixir phase. For either larger or smaller values of $R_c/\sigma$, the elixir phase gradually shrinks until it eventually disappears, as displayed in Figure \ref{fig:fig6}. In the three-dimensional parameter space, $\{1-b/\sigma,\sigma_{sc}/\sigma,R_c/\sigma\}$, the elixir phase is centered at the values of parameters found in proteins, and has a lemon shape with its two ends forming a meeting point for the helix, the beta and the globular phases for large $R_c/\sigma$, and a somewhat extended line \footnote{The origin and the explanation of this line will be discussed elsewhere.} for the helix, the beta and the coil phases for small $R_c/\sigma$. In essence, the elixir phase is an extended co-existence region with a degenerate ground state. 

We note that, while helices and beta strands have the correct geometries within the elixir phase (and not outside it), our simplified model has shortcomings in capturing some of the details of real protein structures. One is the out-of-phase arrangement between parallel strands of a $\beta$-sheet in contrast with the in-phase arrangement in the $\beta$-sheet of real proteins. Another is the difference in the number of residues per turn: 3.6 in real helices and 4 in elixir helices. Both discrepancies can be cured by introducing a binormal-binormal interaction in the model between the Frenet reference frames of main chain spheres in spatial proximity with each other.  We stress that our simple model is adequate for identifying the elixir phase and our goal here is not to mimic the glorious complexities of protein structures by adding more details.

Form determines function for proteins. Furthermore, many globular proteins serve as extraordinarily powerful machines and catalysts. Bernal \cite{Bernal39} wrote, {\it Any effective picture of protein structure must provide at the same time for the common character of all proteins as exemplified by their many chemical and physical similarities, and for the highly specific nature of each protein type.} The amazing common characteristics of proteins along with our observation of the elixir phase in a simple chain model emboldens us to make a constructive hypothesis that protein native state structures may lie in a phase of matter, which confers these properties and the attendant advantages on them.  Our model does not rely on quantum chemistry except in an emergent sense. It does not incorporate hydrogen bonds, the approximate planarity of the peptide bond, the heterogeneity of the side chains, the molecular nature of the rich variety of amino acids and the important role played by the solvent water molecules. A consequence of our hypothesis is that the menu of protein native state structures is determined not by chemistry but rather by general considerations of geometry and the absence of spurious symmetries. The role of the sequence then would be to choose its native fold from this menu in a harmonious manner accounting for the highly specific nature of each protein type. Unlike an earlier study of Zhang et al. \cite{Zhang06}, here we do not consider either an all-atom description of a protein or incorporate hydrogen bonds. It would be interesting to study the coil phase to elucidate the effective scaling exponent of the algebraic dependence of the radius of gyration on chain length\cite{Riback17} to assess whether the unfolded state of foldable sequences is expanded to suppress misfolding and aggregation.


\section{Discussion}
\label{sec:discussion}
We find that the library of observed protein folds is a consequence of basic physical law rather than deriving from the considerations of sequence diversity and chemistry. The finiteness of the number of folds allows for conservation of the geometrical lock-key mechanism central to ligand binding, protein-protein interactions and functionality even as sequence evolution takes place. Our work suggests that Darwinian evolution shapes sequences and functionalities within the constraints of the immutable Platonic folds in the elixir phase. While many aspects of living systems here on earth can be understood within the framework of the theory of evolution, there are important constraints, enormous simplicities, as well as huge advantages deriving from the fact that basic physical law governs living matter. 

Physical systems in the vicinity of a phase transition exhibit exquisite sensitivity to small perturbations of the right kind. Being poised in the vicinity of transitions to other phases affords significant advantages to proteins accounting for their power as molecular machines.
We wish to emphasize that while out model is motivated by symmetry considerations and is generic and simple, it is directly inspired by protein chemistry. In the protein context, the absence of spurious symmetries arise from the presence of side chains, the self-attraction is mediated by water and the effect of hydrophobicity and the overlap of neighboring main chain spheres is a direct consequence of quantum chemistry and covalent bonding. What is truly remarkable is the stunning convergence (or fit) of the chemistry into the simple model we have studied. A corollary of our study is that protein-like behavior ought to be realizable in multiple unrelated ways opening a potential frontier in the creation of nifty machines and even artificial life.
In summary, we find a new phase of matter, which has many attributes similar to those found in proteins. The elixir phase is observed over a range of temperatures. The ground state structures are zero temperature structures. The structures are not the same as protein native state structures. Rather, the elixir phase has certain features that make it a truly novel phase: it is a phase that is observed for a moderate length chain; the structures within it are comprised of helices and sheets; the phase is nestled between other phases and this confers exquisite sensitivity to this phase, which, in turn, makes it suitable for housing machines; the phase exhibits multiple non-trivially connected ground states allowing for switching between them; the ground states are geometrical structures allowing for a lock-key relationship that underlies life; the phase is observed with just self-avoidance and a simple self-attraction regardless of how they arise — the self avoidance does not have spurious symmetries and the self attraction can be mediated by water molecules and hydrophobicity; hydrogen bonds and all-important chemical details including amino acid specificity are not considered but it is quite remarkable how well they fit into the generic scheme; the notion of a phase means that there may be a plethora of ways in which a system can be generated that is housed within the phase; our work therefore transcends the glories of proteins but includes it as a special case.

While, to the best of our knowledge, the elixir phase has not been identified in any previous studies, the model studied here, but with parameters poising the system outside the elixir phase, has been found to exhibit the characteristics of a machine by spontaneously switching between two distinct geometries, a single helix and a double helix, merely because of thermal fluctuations\cite{Banavar09}. A consequence of the existence of the elixir phase is that it can be exploited for the creation of nifty artificially made interacting nano machines. 

The elixir phase is distinct from conventional phases of matter in that the variety of ground state structures is geometry-based and occurs for finite size systems. Our model may be thought of as a generalization of the liquid crystal phase in a chain topology. Just like liquid crystals, the elixir phase structures are stable yet sensitive. Unlike liquid crystals, which occur at non-zero temperatures, the sculptures in the elixir phase are ground states. The elixir phase structures are neither relatively open structures (as in the coil phase) nor are they maximally compact (as in the globular phase). They lie within a marginally compact phase. The ground states in the elixir phase are neither non-degenerate (as in the helix phase) nor do they have a huge degeneracy (as in the globular and especially the coil phases). The elixir phase has an intermediate degeneracy. These 'Goldilocks'-like just-right characteristics make the elixir phase an attractive candidate for facilitating functionality.

The elixir phase is distinct from what is known as the \textit{molten globule} that is believed to be an intermediate state of the folding process \cite{Baldwin13}. Rather, the elixir phase ground states are the counterparts of the native states of globular proteins. 

The elixir phase has characteristics vaguely similar to spin glasses \cite{Edwards75}. Both a spin glass and the elixir phase have energy landscapes with multiple stable minima with barriers between them. This property of spin glasses arises from frustration in conflicting interactions and has been invoked to model content addressable memories \cite{Hopfield82} and prebiotic evolution \cite{Anderson83,Stein84}. The protein free energy landscape has been described to be minimally frustrated leading to a folding funnel geometry \cite{Onuchic97,Dill08,Wales06}. Of course, in the extreme limit of an unfrustrated system, one obtains, in standard spin models, a unique ground state along with its symmetric partners (e.g. all up spins or all down spins for an Ising ferromagnet). The elixir phase is novel in that there is no conventional frustration  -- rather, there are geometrical constraints that allow for distinct chain conformations (multiple ground states) to avail of roughly the same attractive energy; the system is not infinitely large; and the ground states are modular, geometry-based structures related by the distinct topologies of the assembled secondary structures. In fact, the presence of these ground state folds in a homopolymer model suggests that the top of the folding funnel is engineered with no sequence information into many broad basins and the latter part of dynamical folding entails the harmonious fitting of a minimally frustrated sequence into its native state basin. Our picture leads to additional simplicities for understanding and exploring the relative ease of the folding dynamics of globular proteins \cite{Englander17,Baldwin17}. It would be interesting to look for experimental verification of the existence of the elixir phase in colloidal systems where exquisite control can be achieved \cite{Zerrouki08,Wang12}, as well as possible refinements of the present approach.

The elixir phase is the  ground state of our highly simplified model.
Thus it is the analog of the native state in real proteins. However it can also be thought of
as an \textit{approximate} description of the molten globule in real proteins \cite{Baldwin13}, that is an unrefined
conformation within which secondary structures have still not been fully refined.
Of course, this lack of refinement of secondary structures is the result of an incomplete model, e.g. due to the absence of hydrogen bonds. Nevertheless, the approximate structures can be exploited by using the ground states of the elixir phase as a computationally
efficient starting point for a more refined calculation  \cite{Shaw10,Ovchinnikov17} that introduces
additional chemical details. It would be extremely interesting to combine the present strategy with complementary coarse-grained approaches that have been proposed recently in the literature \cite{Hoang04,Craig06,Coluzza11,Bore18} in order to shed new light on the nature of protein folding pathway \cite{Englander17,Baldwin17}.

A cell is not just a container of ordinary molecules - rather, it consists of incredibly powerful interacting molecular machines that orchestrate life. Proteins are amazing catalysts that speed up reactions by many orders of magnitude and carry out many of the functions of a living cell. They are essential ingredients of life. Darwin wrote about the origins of life: {\it But if we could conceive in some warm little pond with all sorts of ammonia $\&$ phosphoric salts,--light, heat, electricity.. present, that a protein compound was chemically formed, ready to undergo still more complex changes...} Our work suggests that nature may have stumbled upon the elixir phase here on earth and eventually this resulted in life as we know it. This opens up the intriguing possibility that life elsewhere in our cosmos could well have a very different chemical basis with some of its vital molecules populating the same elixir phase of matter.

\acknowledgments{We are indebted to Brian Matthews, Flavio Romano, George Rose, Francesco Sciortino, and Pete Von Hippel for useful discussions. This work was supported by MIUR PRIN-COFIN2010-2011 (contract 2010LKE4CC). The use of the SCSCF multiprocessor cluster at  the Universit\`{a} Ca' Foscari Venezia is gratefully acknowledged. T.X.H. acknowledges support from Vietnam National Foundation for Science and Technology Development (NAFOSTED) under Grant No. 103.01-2016.61}


\end{document}